# Resolving Local Electrochemistry at the Nanoscale via Electrochemical Strain Microscopy: Modeling and Experiments

Ahmad Eshghinejad [1,*], Ehsan Nasr Esfahani [1,*], Chihou Lei [2,*], and Jiangyu Li [1,3,†]

1. Department of Mechanical Engineering, University of Washington, Seattle, WA 98195-2600, USA
2. Department of Aerospace and Mechanical Engineering, Saint Louis University, Saint Louis, Missouri, 63103-1110, USA
3. Shenzhen Key Laboratory of Nanobiomechanics, Shenzhen Institutes of Advanced Technology, Chinese Academy of Sciences, Shenzhen 518055, Guangdong, China

**Abstract**

Electrochemistry is the underlying mechanism in a variety of energy conversion and storage systems, and it is well known that the composition, structure, and properties of electrochemical materials near active interfaces often deviates substantially and inhomogeneously from the bulk properties. A universal challenge facing the development of electrochemical systems is our lack of understanding of physical and chemical rates at local length scales, and the recently developed electrochemical strain microscopy (ESM) provides a promising method to probe crucial local information regarding the underlying electrochemical mechanisms. Here we develop a computational model that couples mechanics and electrochemistry relevant for ESM experiments, with the goal to enable quantitative analysis of electrochemical processes underneath a charged scanning probe. We show that the model captures the essence of a number of different ESM experiments, making it possible to de-convolute local ionic concentration and diffusivity via combined ESM mapping, spectroscopy, and relaxation studies. Through the combination of ESM experiments and computations, it is thus possible to obtain deep insight into the local electrochemistry at the nanoscale.

---

[*] These authors contributed equally to the work.

[†] Author to whom the correspondence should be addressed; email: jjli@uw.edu.

**Introduction**

Electrochemistry is the underlying mechanism in a variety of energy conversion and storage systems including Li-ion batteries [1,2], fuel cells [3,4] and solar cells [5,6], and the electrochemical processes in these systems, particularly the flows of electrons and ions, are intimately coupled to mechanics [7]. For example, the electrochemically driven diffusion of ions results in volumetric expansion and contraction in materials, leading to the so-called *Vegard strain* [8,9] that impacts fatigue and failure of the electrodes and solid-state electrolytes [10,11]. Furthermore, the stress associated with the Vegard strain shifts the thermodynamics and kinetics of ionic diffusion, and thus affects the capacity and rate performance of batteries as well [12,13]. While the coupling between ionic diffusion and mechanics often unfavorably affects the performance of an electrochemical system, it also provides a valuable tool to probe the local electrochemical activities, as demonstrated by the recently developed scanning thermo-ionic microscopy (STIM) [14,15] and electrochemical strain microscopy (ESM) [16,17].

It is well known that the composition, structure, and properties of electrochemical materials near active interfaces often deviates substantially and inhomogeneously from the bulk properties [18–20], and a universal challenge facing the development of electrochemical systems is our lack of understanding of physical and chemical rates at local length scales. Conventional electrochemical characterization techniques are based on the measurement of current, and thus are very difficult to scale down to nanometer regime. Electrochemical Vegard strain, on the other hand, provides an alternative imaging mechanism that promises high sensitivity and spatial resolution. For instance, Tian *et al*. [21] used atomic force microscopy (AFM) to monitor the topography evolution of a lithium ion battery electrode upon charging and discharging, while Balke and Kalinin *et al*. [16,17] proposed ESM to investigate local ionic activities excited by a charged scanning probe. Recently we have also developed STIM to excite ionic fluctuation driven by local thermal stress oscillation instead of electrical potential, and probe the induced dynamic Vegard strain accordingly [14,15]. Because of their high sensitivity and nanoscale resolution, ESM and STIM have the potential to probe crucial information regarding the underlying electrochemical mechanisms at the most relevant length scale, such as ionic transport [2,18], interfacial chemistry and charge transfer [22–24], and they have been applied successfully to study a variety of electrochemically active materials in the past few years [25–29].

While considerable insights have been learned from ESM and STIM studies, these experiments remain largely qualitative in nature and the data are rather challenging to analyze and interpret [30]. In fact, it is rather difficult to draw key electrochemical parameters such as local ionic concentration and diffusivity from the experimental data. Furthermore, Chen *et al*. [31] showed that electrostatic interactions and other electromechanical mechanisms often interfere with ESM response in ionic materials, while Yu *et al.* showed that both Vegard and plausible non-Vegard strains contribute to the ESM signals in fresh and aged $LiMn_2O_4$ battery cathodes [32,33]. Thus it is important to understand the characteristics of ESM response arising from Vegard strain, so that different microscopic mechanisms responsible for the observed phenomena can be assessed and differentiated. In this regard, Morozovska *et al*. [34] have proposed a semi-analytical model for ESM analysis [35], focusing on frequency dependence of ESM signals. Such analysis was further extended to nonlinear regime [36] with flexoelectricity incorporated [37], though these two studies concern primarily the distribution of electrochemical and mechanical fields induced by a charged probe that is fixed at origin. Here we seek to develop a realistic computational model that couples mechanics and electrochemistry relevant for ESM experiments, focusing instead on how ESM signals depend on material parameters such as diffusivity and ionic concentration. Furthermore, we are interested in the electromechanical displacement of the heterogeneous sample underneath the scanning (instead of fixed) probe, so that the results can be directly compared with the experimental ESM mapping. The ultimate goal is to enable quantitative analysis of electrochemical processes underneath a charged scanning probe, so that experimental protocols for measuring local ionic concentration and diffusivity can be developed. In this regard, we note some previous success in time and voltage spectroscopies of ESM experiment and modeling that resulted in estimated local diffusivity [33,38]. While this particular model is developed for ESM, it can be extended for STIM analysis if we incorporate heat transfer process underneath of a thermal probe [39].

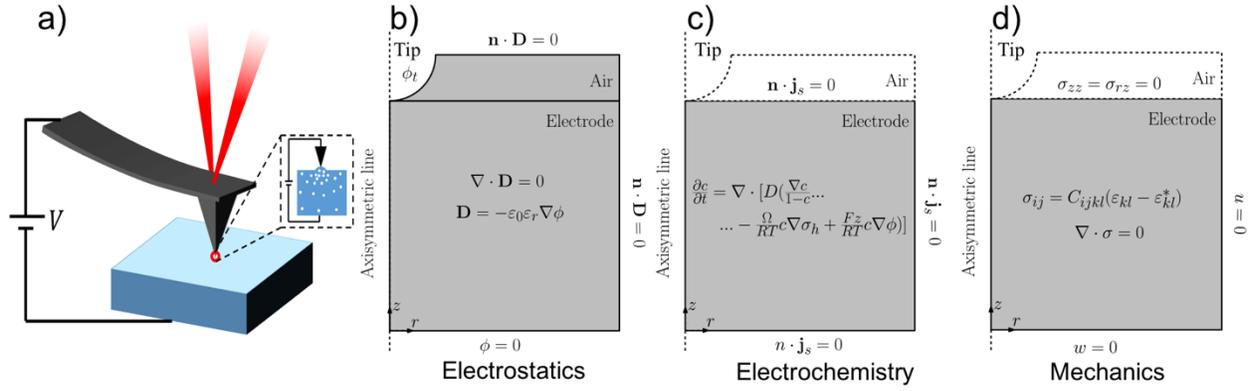

**Fig. 1** The configuration of computational model of ESM; (a) schematics of ESM, wherein ionic oscillation is excited by a charged probe, resulting in Vegard strain; (b-d) the axisymmetric computational domains, boundary conditions, and governing equations for (b) electrostatics, (c) electrochemistry, and (d) mechanics in the specimen underneath the charged probe.

**Modeling Framework**

In an ESM experiment, as schematically shown in Fig. 1(a), a time variant bias is applied to the conductive probe, inducing a change in the electrochemical potential of the sample. This in turn triggers ionic oscillation underneath the probe, resulting in an oscillating *Vegard strain* [8,9] and the corresponding surface vibration that can be measured by the AFM photodiode through the cantilever deflection signal. As such, the problem involves coupled electrostatics, electrochemistry and mechanics. The bias-induced electric potential $\phi$ can be solved from the Laplace equation with electric neutrality,

$$\nabla \cdot \boldsymbol{D} = \boldsymbol{0}, \qquad \boldsymbol{D} = -\varepsilon_0 \varepsilon_r \nabla \phi, \tag{1}$$

where $\boldsymbol{D}$ is the electric displacement, and $\varepsilon_0$ and $\varepsilon_r$ are the vacuum permittivity and the relative permittivity of the probed medium. As shown in Fig. 1(b), we assume a spherical probe tip and a cylindrical sample with radius ratio of 1/10, and thus the problem is axisymmetric. The potential is specified on the conductive probe, and the specimen is grounded via a bottom substrate, as in a typical ESM experiment. The circumferential boundary of the specimen is assumed to be charge-free. The bias imposed by the probe shifts the electrochemical potential $\mu$ of ionic species (with the charge *z*) in the material [40],

$$\mu = RT \ln\left(\frac{c}{1-c}\right) - \Omega\sigma_h + Fz\phi, \tag{2}$$

where $R, T, F$, and $\sigma_h$ are the gas constant, temperature, Faraday constant, and the hydrostatic stress, respectively, $\Omega$ is the partial molar volume, and $c = \frac{\hat{c}}{c_{max}}$ is the dimensionless concentration normalized with respect to $c_{max}$, the maximum allowable ionic concentration in the material. In Eq. (2), the first term arises from entropic force corresponding to the energy change associated with introducing the ionic species into a fixed set of host sites [41], the second term arises from mechanical potential associated with introducing ions into a solid host with the partial molar volume $\Omega$ [12,42], and the last term is due to the interaction between ions and the externally applied electric potential $\phi$. The inhomogeneous electrochemical potential distribution thus results in ionic flux $\boldsymbol{j}$ and redistribution of ionic concentration governed by Fick's law,

$$\frac{\partial c}{\partial t} = -\nabla \cdot \boldsymbol{j} = \nabla \cdot \left[D\left(\frac{\nabla c}{1-c} - \frac{\Omega}{RT}c\nabla\sigma_h + \frac{Fz}{RT}c\nabla\phi\right)\right], \tag{3}$$

where $t$ and $D$ are the time and diffusivity constant, respectively. As shown in Fig. 1(c), the system is assumed to be closed with no electrochemical reactions and mass transfers at the boundaries. The ionic redistribution within the sample, however, results in Vegard strain $\boldsymbol{\varepsilon}^*$ and the corresponding stress $\boldsymbol{\sigma}$ that can be solved through the elastic equilibrium equation [8,9],

$$\boldsymbol{\varepsilon}^* = \frac{\Omega}{3}c_{max}c\mathbf{I}, \qquad \boldsymbol{\sigma} = \boldsymbol{C}(\nabla\boldsymbol{u} - \boldsymbol{\varepsilon}^*), \qquad \nabla\cdot\boldsymbol{\sigma} = 0, \tag{4}$$

where $\boldsymbol{I}$ is the second-order unit tensor, $\boldsymbol{C}$ is the stiffness tensor of the material and $\boldsymbol{u}$ is the resulted displacement, which is responsible for the observed displacement measured through cantilever deflection. As shown in Fig. 1(d), the top surface is assumed traction free, and no vertical ($w$) and horizontal ($u$) displacements are allowed on the bottom and side surfaces, respectively. This is the essence of our model, which was implemented in COMSOL Multiphysics package. Note that we ignored the topographical variation of the sample, and considered the surface to be smooth for the time being. In principle, it can be incorporated into the COMSOL simulation without much difficulty.

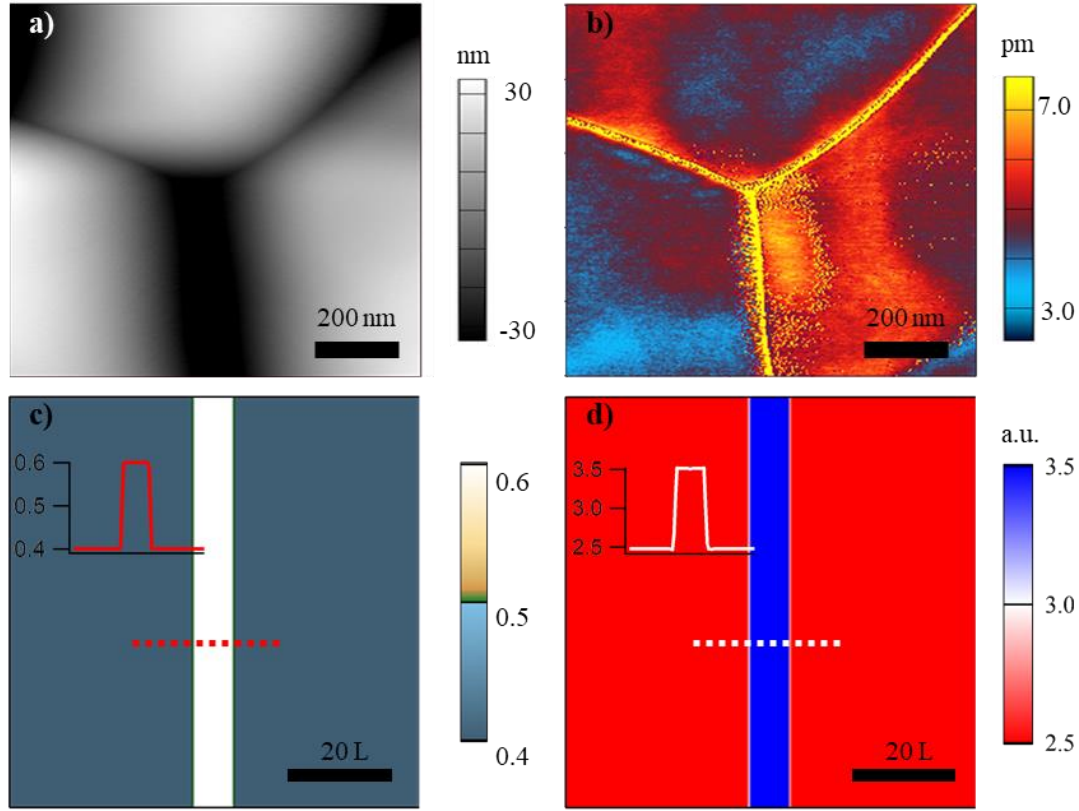

**Fig. 2** ESM mapping from experiment and simulation; (a) Sm-doped ceria topography and (b) ESM amplitude mappings acquired from ESM experiment; (c) pre-imposed distribution of dimensionless baseline ionic concentration $c_0$ used in the simulation and (d) the corresponding ESM amplitude mapping simulated. The simulation size is equal to $80L$ over on $200\times200$ pixels, with $L$ being probe radius.

**Results and Discussions**

*ESM Amplitude*

We first consider ESM mapping, i.e. the distribution of sample surface displacement underneath the scanning probe excited by the AC voltage. One such an ESM mapping is shown in Fig. 2(b), where a triple junction region of a nanocrystalline Sm-doped ceria sample was scanned, as revealed by its topography in Fig. 2(a). An inhomogeneous ESM amplitude distribution is observed, particularly at the grain boundaries, where the response appears to be enhanced, as previously reported by Chen *et al.* [43]. What does such distribution mean then, and what kind of insight can we learn from it? It is well known that the nanocrystalline ceria demonstrates orders of magnitude higher ionic conductivity than its bulk counterpart [44], which was theorized to be caused by accumulation of space charges at the grain boundaries [45,46]. Chen *et al.* [43] invoked the potential-induced changes of small polaron concentration in the space-charge regions to explain their experimental ESM data. In order to rationalize such a theory and the corresponding experimental observation, ESM simulations were carried out, which treats polaron in a similar way as mobile ions with a pre-imposed base-line concentration shown in Fig. 2(c). The higher polaron concentration at an idealized grain boundary was specified based on the proposed theory [45,46]. The corresponding ESM amplitude mapping simulated under the charged probe reveals a clear correspondence between the ESM amplitude and ionic concentration as shown in Fig. 2(d), i.e. the higher ionic concentration leads to an enhanced ESM response.

It should be noted that the experimentally measured ESM amplitude is resonance-enhanced, and thus sensitive to the variations of the tip-sample interaction. While the rough topography contrast between grain and grain boundaries could result in different contact stiffness, this effect has been addressed to certain extent by using dual amplitude resonance tracking (DART) technique [47] and corrected by the simple harmonic oscillator (SHO) model [48]. In other words, our experimentally measured ESM amplitude is intrinsic, with resonance enhancement and variation of contact stiffness corrected.

These results suggest that the ESM amplitude correlates with ionic concentration, and indeed, this can be appreciated from a zero-order Taylor expansion of Eq. (3) around a baseline concentration $c_0$,

$$\delta c = D c_0 \nabla \cdot \left( -\frac{\Omega}{RT} \nabla \sigma_h + \frac{Fz}{RT} \nabla \phi \right), \tag{5}$$

which clearly indicates that the quasi-static changes in the instantaneous ionic fluctuation, and thus the ESM amplitude, scale with the baseline concentration $c_0$ and diffusivity $D$.

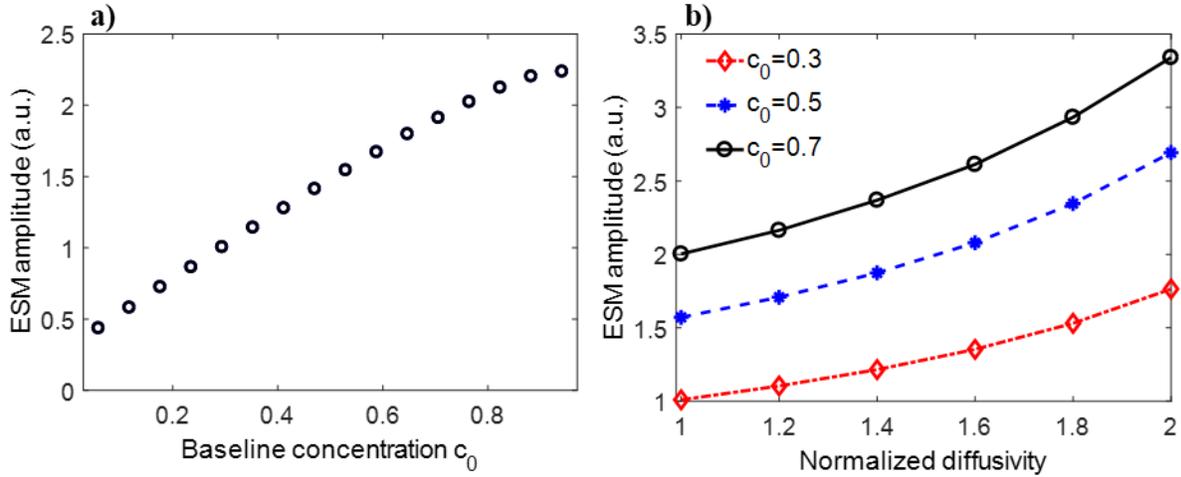

**Fig. 3** Effects of the ionic concentration and diffusivity on the ESM response; simulations of ESM amplitude as a function of (a) baseline concentration and (b) the normalized diffusivity.

In order to verify this analysis, we also calculate the ESM amplitude a function of the baseline ionic concentration $c_0$ that is assumed to be uniformly distributed for simplicity, as shown in Fig. 3(a), revealing good linear correlation (except at high concentration end), as predicted. If we fix the baseline concentration $c_0$ while vary the ionic diffusivity $D$, as shown in Fig. 3(b), it is then observed that the simulated ESM amplitude increases as the diffusivity increases, and the deviation from the linear relationship is not significant. It is worth mentioning that the experimentally measured ESM amplitude is a convolution of both baseline concentration $c_0$ and diffusivity $D$, which is considered to represent electrochemical activities. Therefore, it is concluded that the higher ESM amplitude observed in the grain boundaries of Sm-doped ceria corresponds to the higher electrochemical activities, consistent with the proposed accumulation of space charges at the grain boundaries [45,46].

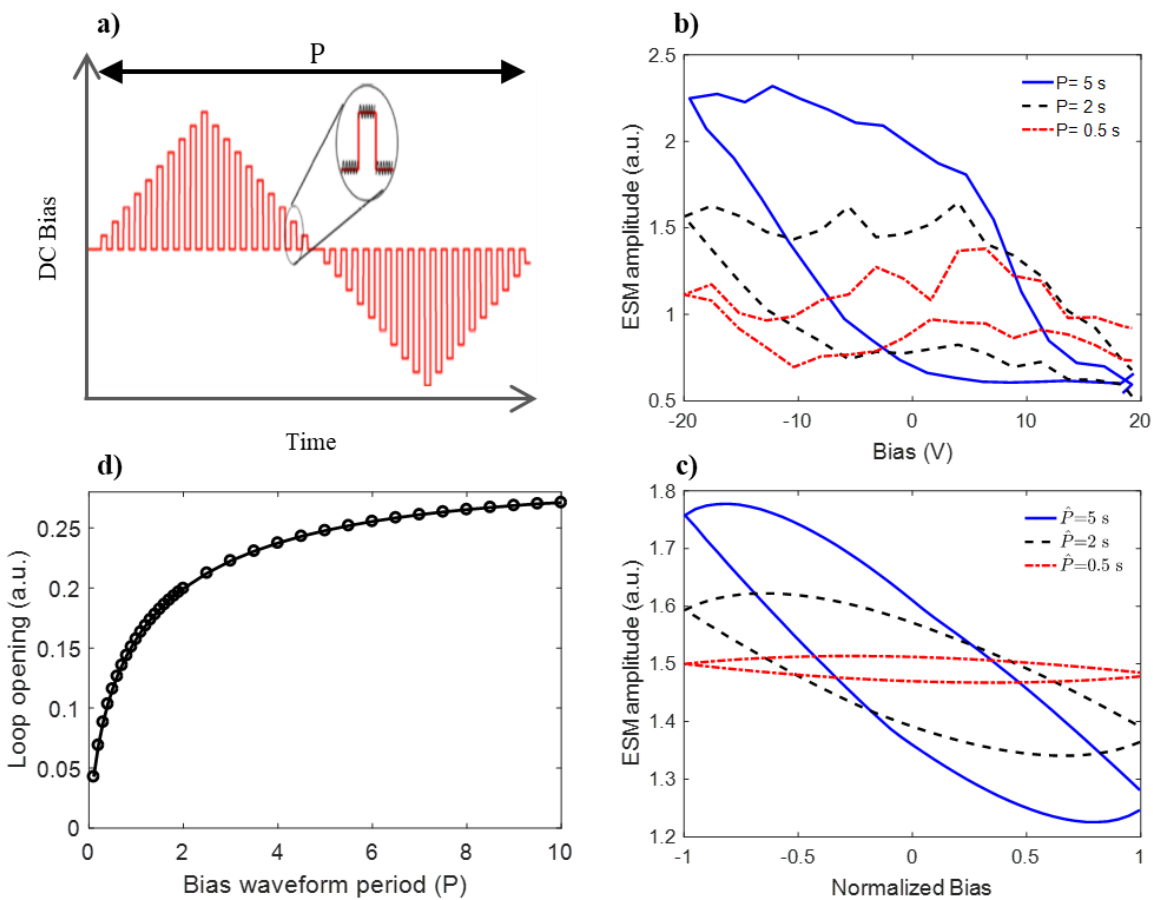

**Fig. 4** ESM spectroscopy study of a ceria sample; (a) the bias waveform applied to the probe; (b) measured and (c) simulated hysteresis loops between the applied ESM amplitude and bias; (d) simulated loop opening area versus bias period. Each bias period contains 35 pulses with equal duration for off- and on-state.

## *ESM Bias Spectroscopy*

The analysis on ESM amplitude raises an important question - how can we decouple the effects of the local ionic concentration and diffusivity in ESM experiments? Spectroscopy experiments provide one possible means, where a triangular DC bias waveform, as shown in Fig. 4(a), is applied to the probe to slowly change the baseline concentration while the superposed AC bias induces fast ionic fluctuations that allows for instantaneous ESM measurements. In order to minimize electrostatic interactions between the charged probe and sample, the instantaneous ESM amplitudes are measured and averaged in short periods when the DC bias is stepped back to zero (off-state). The off-state experimental ESM amplitudes as a function of DC bias with three different bias periods *P* are measured on a LiFePO$_4$ sample, as shown in Fig. 4(b), which clearly reveals different shapes of hysteresis loops. The observed ESM amplitude, which is directly related to the ionic concentration, is higher under the negative biases that attract the positively charged species, consistent with our previous analysis [28,43]. Both hysteric behavior and reduction of the ESM amplitude as a function of the DC bias are captured well in our simulation, as shown in Fig. 4(c).

Note that these amplitude-bias hysteresis loops are distinct from butterfly loops in ferroelectric materials resulted from polarization switching, indicating that the polarity of deformation, arising from Vegard strain, does not change under opposite DC bias. They should not be confused with phase-bias hysteresis loop in piezoresponse force microscopy (PFM) as well [30,49,50], as noted by Chen *et al.* [31], and it can be used to distinguish electrochemical strain from piezoelectric strain.

The shape of ESM loop is determined by the kinetics and dynamics of electrochemical reaction and thus the loops are highly rate-dependent [33,51]. Importantly, in both simulations and experiments, it is observed that applying the DC bias with longer period results in higher ESM amplitude and larger hysteresis loop area. The hysteresis loops open up when the waveform period *P* is increased, since it allows longer time and higher extent of ionic redistribution over a longer range. Computationally, such a longer period is equivalent to a higher diffusivity (under constant time) since we use normalized time defined as $\hat{t} = \frac{tD}{L^2}$, as detailed in the Methods, and thus it could provide us a mean to analyze the dynamics associated with local ionic redistribution. To this end, we evaluated the loop opening, defined as the difference in amplitudes at zero bias, versus the period (or analogously diffusivity), as shown in Fig. 4(d), wherein it is observed that the opening increases rapidly as the period (diffusivity) increases initially, before it reaches a plateau that corresponds to the equilibrium ionic distribution underneath the biased probe.

To investigate the effects of baseline concentration and diffusivity on the hysteresis loops, ESM spectroscopy studies were carried out on the grain and grain boundaries in a triple junction area of LiFePO$_4$, as shown in Fig. 5(a), where the ESM mapping is overlaid on 3D topography, exhibiting again enhanced response at grain boundaries. The hysteresis loops measured within grain and at the grain boundary are shown in Fig. 5(b), for which both loops were averaged over three points to show the general contrast. It is observed that the grain boundary not only exhibits enhanced ESM amplitude, but also larger loop opening compared to point probed within the grain. This is consistent with the experimental results on Si electrode reported previously [25]. Does the difference in loop opening arise from the diffusivity, or ionic concentration? To answer this question, we simulated spectroscopy experiments across the grain boundary with specified concentration distribution as shown in Fig. 2(c), where higher ionic concentration is specified at the grain boundary. The simulated hysteresis loops are shown in in Fig. 5(c), revealing slightly larger loop opening at grain boundaries, but to much less extent compared to experimental measured one. This suggests that the main contribution to the observed loop opening arises from different diffusivity, as exhibited in Fig. 4, instead of the ionic concentration, and this points toward a method for estimating local diffusivity.

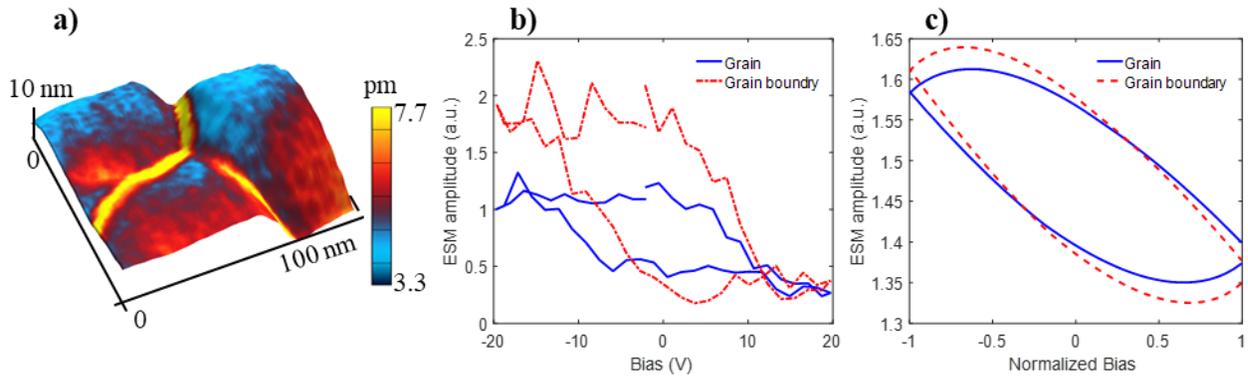

**Fig. 5** ESM spectroscopy studies from (a,b) experiments and (c) simulation within a grain and at a grain boundary; (a) ESM amplitude mapping overlaid on 3D topography; (b) hysteresis loops measured within a grain and on a grain boundary; (c) simulated hysteresis loops within a grain and on a grain boundary.

### *Relaxation Study*

While the spectroscopy hysteresis loop opening correlates with the diffusivity, as revealed by Fig. 4(d), the local diffusivity is better investigated by relaxation study under a stepwise DC bias, as shown in Fig. 6, also known as time spectroscopy. This DC bias modifies the local ionic concentration underneath the probe, and when it is removed, the response will relax back to the level corresponding to the baseline concentration. It is thus possible to estimate the diffusivity using the time constant measured from the relaxation curve. Such relaxation behavior is evident from experimental data measured on $LiFePO_4$ [26] shown in Fig. 6(a). The positive DC reduces local concentration of lithium ions, and thus the ESM response decreases, and then relaxes back to the baseline response. The negative DC bias, on the other hand, increases the local ionic concentration and thus ESM response, which also drops when the DC bias is removed. Similar trends have been reported by Amanieu *et al.* [38] as well, and the behavior are clearly captured in the simulation shown in Fig. 6(b).

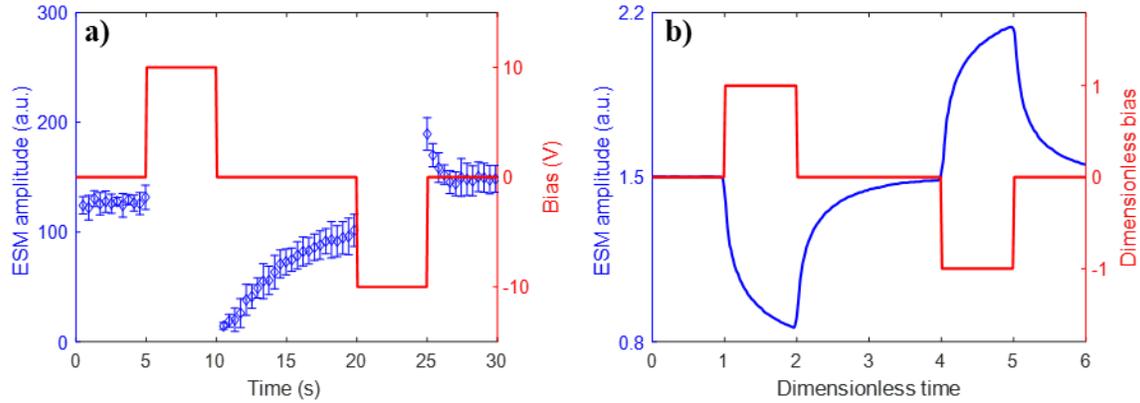

**Fig. 6** ESM relaxation studies on a LiFePO$_4$ sample; (a) experimental and (b) simulated ESM amplitude versus time measured on LiFePO$_4$ under a stepwise bias [26]; the left axis indicates the ESM amplitude and the right axis indicates the applied bias.

To better understand the dynamic behavior, we examine the variation of local ionic concentration versus time in one-dimension, after the DC bias is dropped and the effect of stress is ignored, such that

$$\frac{\partial c}{\partial t} = \nabla \cdot (D \nabla c) = D \nabla^2 c = D \frac{\partial^2 c}{\partial x^2}. \tag{6}$$

The general form of the solution for this diffusion equation subjected to Neumann boundary conditions for a closed system can be obtained via the method of separation of variables:

$$c(x,t) = c + \sum_{m=1}^{\infty} c_m e^{-D\pi^2 \left(\frac{m^2}{l^2}\right) t} \cos\left(\frac{m}{l} \pi x\right), \tag{7}$$

where $l$ is the dimensions of the domain, and $c_m$ are constants determined from the initial condition using Fourier Series:

$$c_m = \frac{2}{l} \int_0^l c(x,0) \cos\left(\frac{m}{l} \pi x\right) dx, \quad \forall m = 1, 2, \ldots,$$

In particular, the steady state concentration, as $t \to \infty$, is a uniform distribution:

$$c(x, t \to \infty) = c_0 = \frac{1}{l}\int_0^l c(x, 0)dx = \langle c(x, 0)\rangle_{ave} = \langle c(x, t)\rangle_{ave}.$$

Importantly, the time dependent part of the solution indicates an exponential decay, as demonstrated by the term:

$$e^{-D\pi^2\left(\frac{m^2}{l^2}\right)t} \sim e^{-\frac{t}{\tau}}, \tag{8}$$

Where the time constant

$$\tau = \tau_m = \frac{1}{D\pi^2\left(\frac{m^2}{l^2}\right)} \tag{9}$$

is inversely proportional to the diffusivity $D$, and thus can be used to estimate it. The analysis can be easily generalized to two- or three-dimensional models, and the essence of the conclusion remains.

**Concluding Remarks**

In this study, we have developed a coupled modeling framework to compute the electrochemical processes underneath a charged probe, and show that it captures the essence of a number of different ESM experiments. We demonstrate that the ESM response correlates with both local ionic concentration and diffusivity, while spectroscopy hysteresis behavior and relaxation time constant are mostly governed by local diffusivity, in good agreement with experimental observations. Thus, through the combination of ESM mapping and point-wise voltage and time spectroscopies, it is possible to de-convolute local ionic concentration and diffusivity in ESM experiments, offering deep insight into local electrochemistry with high spatial resolution and sensitivity. The model can be extended to other techniques based on a scanning probe as well, particularly STIM that excite the local ionic activities through thermal stress instead of electric field, making it easier for in-operando imaging by minimizing electric interference from the global voltage/current perturbation.

## Methods

### *ESM experiments*

The ESM experiments were conducted on Asylum Research MFP-3D AFM using NanoSensors PPP-EFM probes with Ptlr5 metallic coating having tip radius of 25 nm and nominal resonance frequency of 70 kHz in air. The drive amplitude for the ESM mapping and spectroscopy studies was 3 and 1 V, respectively. The dual AC resonance tracking of the tip-sample contact was used to enhance the signal to noise ratio [52]. The spectroscopy experiments were repeated in 3 loops (the waveform shown in Fig. 4(a) was cycled 3 times) over 3 points and under three different periods, and the intrinsic amplitude was found through SHO fitting, with the averaged results presented. Each period for $P$=0.5, 2 and 5 s contains 35 short pulses with equal durations for off- and on-states, resulting in pulse durations of 7.1 ms, 28.6 ms, and 71.4 ms, respectively.

### *FEM simulation*

All the simulations were carried out using COMSOL Multiphysics package. To facilitate analysis, the following normalizations were adopted,

$$\hat{x} = \frac{x}{L}, \qquad \hat{t} = \frac{tD}{L^2}, \qquad \hat{\phi} = \frac{\phi Fz}{RT}, \qquad \hat{\sigma}_h = \frac{\sigma_h \Omega}{RT},$$

where $L$ is the reference length, taken as the tip radius. The electrostatics and electrochemistry governing equations were implemented into the General PDE module and the mechanics was implemented into the Solid Mechanics module. The simulations were conducted using axisymmetric models except for the ESM mappings, which was performed using a 3D model. The following materials constants were used in the simulation [41,53].

Table 1 Material properties used for simulations of ceria and LiFePO$_4$ [53,54].

| Material | Modulus of elasticity $E$ | Poisson's ratio $\nu$ | Partial molar volume $\Omega$ | Maximum concentration $c_{max}$ | Electrode permittivity $\varepsilon_r$ |
|---|---|---|---|---|---|
| ceria | 165Gpa | 0.3 | $5.4 \times 10^{-6}$ m$^3$/mol | 18,450 mol/m$^3$ | 10 |
| LiFePO$_4$ | 100Gpa | 0.3 | $3.5 \times 10^{-6}$ m$^3$/mol | 22,900 mol/m$^3$ | 10 |

**Competing Interests**

There are no competing interests.

**Contributions**

JL conceived the project, AE, CL, EE, and JL developed the model, AE implemented the computations and carried out experiment, EE revised the article and assisted the data processing and analysis. AE, EE, and JL wrote the manuscript, and all the authors revised the manuscript.

**Funding**

This material is based in part upon work supported by National Key Research and Development Program of China (2016YFA0201001), National Natural Science Foundation of China (11627801 and 11472236), US National Science Foundation (CBET-1435968), the Leading Talents Program of Guangdong Province (2016LJ06C372), Key Laboratory for Magnetic Resonance and Multimodality Imaging of Guangdong Province (2014B030301013), and the State of Washington through the University of Washington Clean Energy Institute.